\documentclass[prd,floatfix,twocolumn,amsmath,amssymb,nofootinbib]{revtex4}
\usepackage{}
\usepackage{graphicx,color,dcolumn,booktabs,bm}
\usepackage{longtable,lscape}
\usepackage{txfonts}
\usepackage{overpic}
\usepackage{amssymb}
\usepackage{epstopdf}
\usepackage{indentfirst}
\usepackage{feynmf}   
\usepackage{slashed}  
\usepackage{cases}
\usepackage{color}
\usepackage{float}
\usepackage{multirow}
\usepackage{graphicx,color,dcolumn,booktabs,bm}
\usepackage{epsfig,dsfont,amssymb,amsmath,amsfonts,amsbsy,mathrsfs}

\graphicspath{{Figures/}} %
\usepackage[colorlinks, citecolor=blue,anchorcolor=red,menucolor=red, linkcolor=red,filecolor=red,runcolor=red,urlcolor=blue,frenchlinks=red]{hyperref}

\begin{document}

\title{Regge-like relation and a universal description of heavy-light systems}

\author{Kan Chen$^{1,2}$}\email{chenk_10@lzu.edu.cn}
\author{Yubing Dong$^{3,4,5}$}\email{dongyb@ihep.ac.cn}
\author{Xiang Liu$^{1,2}$}\email{xiangliu@lzu.edu.cn}
\author{Qi-Fang L\"u$^{3,6}$}\email{lvqifang@ihep.ac.cn}
\author{Takayuki Matsuki$^{7,8}$}\email{matsuki@tokyo-kasei.ac.jp}

\affiliation{
$^1$School of Physical Science and Technology, Lanzhou University, Lanzhou 730000, China\\
$^2$Research Center for Hadron and CSR Physics, Lanzhou University and Institute of Modern Physics of CAS, Lanzhou 730000, China}
\affiliation{
$^3$Institute of High Energy Physics, CAS, Beijing 100049, China\\
$^4$Theoretical Physics Center for Science Facilities (TPCSF), CAS, China\\
$^5$School of Physical Sciences, University of Chinese Academy of Sciences, Beijing 101408, China
}
\affiliation{
$^6$Synergetic Innovation Center for Quantum Effects and Applications (SICQEA), Hunan Normal University, Changsha 410081, China
}
\affiliation{
$^7$Tokyo Kasei University, 1-18-1 Kaga, Itabashi, Tokyo 173-8602, Japan\\
$^8$Theoretical Research Division, Nishina Center, RIKEN, Wako, Saitama 351-0198, Japan
}

\begin{abstract}

Using the Regge-like formula $(M-m_Q)^2=\pi\sigma L$ between hadron mass $M$ and angular momentum $L$ with a heavy quark mass $m_Q$ and a string tension $\sigma$, we analyze all the heavy-light systems, i.e., $D/D_s/B/B_s$ mesons and charmed and bottom baryons.
Numerical plots are obtained for all the heavy-light mesons of experimental data whose slope becomes nearly equal to 1/2 of that for light hadrons.
Assuming that charmed and bottom baryons consist of one heavy quark and one light cluster of two light quarks (diquark), we apply the formula to all the heavy-light baryons including recently discovered $\Omega_c$'s and find that
these baryons experimentally measured satisfy the above formula.
We predict the average mass values of $B$, $B_s$, $\Lambda_b$,  $\Sigma_c$, $\Xi_c$, and $\Omega_c$ with $L=2$ as 6.01, 6.13, 6.15, 3.05, 3.07, and 3.34 GeV, respectively.
Our results on baryons suggest that these baryons can be safely regarded as heavy quark-light cluster configuration. We also find a universal description for all the heavy-light mesons as well as
baryons, i.e., one unique line is enough to describe both of charmed and bottom heavy-light systems. Our results suggest that instead of mass itself, gluon flux energy is essential to obtain a linear trajectory. Our method gives a straight line for $B_c$ although the curved parent Regge trajectory was suggested before.

\end{abstract}

\maketitle

\section{introduction}\label{sec1}

The nature has chosen the quantum number $^{2S+1}L_J$ to classify light and heavy hadrons including light $u/d/s$ and heavy $c/b$ quarks, respectively. This is true for light hadrons which can be treated nonrelativistically but this also holds for heavy-light mesons which was analytically derived in the former paper \cite{Matsuki:2016hzk} using our semi-relativistic potential model \cite{Matsuki:1997da,Matsuki:2007zza}. Actually this has been noticed and pointed out in a couple of different contexts, e.g., string picture/flux tube model \cite{LaCourse:1988cu,Selem:2006nd}, quantum mechanical derivation of Regge trajectories \cite{Veseli:1996gy}, suppression of $LS$ coupling \cite{Page:2000ij,Riazuddin:2011wr}, empirical rule of degeneracy among states with the same $L$ \cite{Afonin:2007jd,Afonin:2007aa,Afonin:2013hla}, etc.

In the former paper \cite{Matsuki:2016hzk}, we have pointed out that a careful observation of the experimental spectra of heavy-light mesons tells us that heavy-light mesons with the same angular momentum $L$ are almost degenerate. In other words, we have observed that mass differences within a heavy quark spin doublet and between doublets with the same $L$ are very small compared with a mass gap between different multiplets with different $L$, which is nearly equal to the value of $\Lambda_{QCD}\sim 300$ MeV.
This fact is analytically explained by our semirelativistic potential model \cite{Matsuki:1997da} which is proposed to describe heavy-light mesons, spectra and wave functions.
In Refs. \cite{LaCourse:1988cu,Selem:2006nd}, the authors took a simplest string configuration or flux tube picture of a $q\bar q$ meson based on Nambu's idea \cite{Nambu:1974zg} of gluon flux tube picture for a string. They derive a relation between mass squared and angular momentum.
In Ref. \cite{Veseli:1996gy}, the authors also derived the similar relation using quantum mechanics with a Cornell potential model for a meson.
In Refs. \cite{Page:2000ij,Riazuddin:2011wr}, the authors noticed that suppression of $LS$ coupling should occur in the heavy quark symmetry which can be applied to heavy-light mesons.
In Refs. \cite{Afonin:2007jd,Afonin:2007aa,Afonin:2013hla}, carefully checking experimental data, they noticed that light hadrons and vector mesons can be well classified in the angular momentum and also find that one unique line is enough to describe vector mesons including $\phi,\omega$, and $\psi$ mesons which they call {\it a universal description}.

Low lying mesons are classified in terms of $^{2S+1}L_J$, which satisfy the Regge-Chew-Frautch formula $M^2=an+bJ+c$ with constants $a$, $b$, and $c$, principal quantum number $n$, and total angular momentum $J$. Or they satisfy the non-relativistic Regge trajectory for mesons,
\begin{equation}
  L=\alpha' M^2 + \alpha_0, \quad n=\beta M^2+\beta_0, \label{eq:Regge}
\end{equation}
with the Regge slope $\alpha'$ and constants, $\alpha_0,~\beta$, and $\beta_0$. By using the experimental data, the generalized Regge or the Chew-Frautch formula has been studied in detail in Ref. \cite{Afonin:2007jd} which gives the values of constants, $a$, $b$, and $c$. Regge trajectories are still effective in determining spin and parity of newly discovered hadrons.
We can have this type of relation which also holds for heavy-light mesons when one notices that mass gaps between states with different $L$ are nearly equal to $\Lambda_{QCD}$ (see Table 1 in \cite{Matsuki:2016hzk}).

In this paper, instead of using Eq. (\ref{eq:Regge}), as a powerful tool to analyze all the heavy-light systems, we use the formula,
\begin{equation}
  (M-m_Q)^2=\pi\sigma L, \label{eq:ML}
\end{equation}
which was originally derived in Refs. \cite{LaCourse:1988cu,Veseli:1996gy}. In the next section, following Nambu's picture for hadrons in which quarks are connected by a gluon flux tube \cite{Nambu:1974zg}, we will give much simpler derivation of this relation between the heavy-light meson mass and angular momentum.

Then, we apply this formula to heavy-light mesons ($D/D_s/B/B_s$) to obtain linear trajectories in the plane $(M-m_Q)^2$ vs. $L$ using experimental data as well as theoretical values. Next, we will extend this formula to heavy baryons ($\Lambda_c, ~\Lambda_b, ~\Sigma_c, ~\Sigma_b, ~\Xi_c, ~\Xi_b, ~\Omega_c$, and $\Omega_b$) regarding a diquark as a $\bar 3$ color state to see whether it works or not. If this is successful, we can safely say that the heavy-light baryons can be well described by a picture in which one heavy quark couples with a diquark. This is one of the motivations of this paper whether diquark picture holds or not since there are some questions to use this concept for baryons. This ovservation might be checked by the lattice gauge theory without fermion.

We also try to check whether a universal description holds, i.e., whether Regge-like lines for $X_c$ and $X_b$ overlaps or not with $X_Q$ being heavy-light systems. Final section is devoted to conclusions and discussion, especially on the meaning of gluon flux energy, $M-m_Q$, and hadron mass $M$ and on the nonlinearity of the parent Regge trajectories for heavy quark systems \cite{Ebert:2011jc}.

\section{Relation between Mass and Angular Momentum}\label{sec2}

We take Nambu's picture \cite{Nambu:1974zg} of a hadronic string, which consists only of gluons and at both ends of which quarks are attached in the case of mesons.
The authors in Refs. \cite{LaCourse:1988cu,Afonin:2007jd} took the same picture to derive the Regge formula:
\begin{eqnarray}
  M^2 = 2\pi\sigma L, \label{eq:CF1}
\end{eqnarray}
where $M$ is a meson mass, $L$ is an angular momentum and $\sigma$ is a string tension.
Their way of derivation is as follows. See also Refs. \cite{Afonin:2007aa,Afonin:2013hla} for other applications.
Assuming the simple string configuration, we obtain the mass $M$ and angular momentum $L$ given by the following equations,
\begin{eqnarray}
  M &=& 2\int_0^{\ell/2}\frac{\sigma dr}{\sqrt{1-v^2(r)}} = \frac{\pi\sigma \ell}{2}, \label{eq:ML1} \\
  L &=& 2\int_0^{\ell/2}\frac{\sigma rv(r)dr}{\sqrt{1-v^2(r)}}=\frac{\pi\sigma\ell^2}{8},
  \label{eq:ML2}
\end{eqnarray}
where $\ell$ is a length of a string which connects two light quarks at the ends, and $v(r)=2r/\ell$ is the speed of the flux tube at the distance $r$ from the center of rotation.
Equations (\ref{eq:ML1}) and (\ref{eq:ML2}) are obtained by assuming the simplest configuration of a string which connects two quarks at both ends of a string with a speed of light $c$ and rotates around the center of the mass system. Combining these two equations, we arrive at Eq. (\ref{eq:CF1}).

\begin{figure}[htpb]
	\begin{center}
		\includegraphics[scale=0.5]{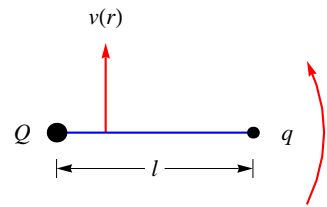}
		\caption{(color online). The schematic diagram for depicting the string connecting heavy and light quarks at the ends and rotating around the heavy quark. A heavy quark is fixed at one point.}\label{stringfig}
	\end{center}
\end{figure}

Let us apply the above idea to a heavy-light meson. In the heavy quark limit, one considers the situation that a heavy quark is fixed at one end and a light quark rotates around a heavy quark with a speed of light $c$ as described in Fig. \ref{stringfig}. Then one obtains the following relation:
\begin{eqnarray}
  M^2 = \pi\sigma L. \label{eq:CF2}
\end{eqnarray}
The right hand side of this equation is just 1/2 of Eq. (\ref{eq:CF1}) and our numerical plot does not fit with this equation.
Because the Nambu's string consists only of gluons, we need to get rid of effects of quark masses. In our case, in the limit of heavy quark effective theory, we should subtract only the heavy quark mass, $m_Q$ with $Q=c,~b$, from $M$. Hence the final expression we should adopt is given by
\begin{eqnarray}
  \left(M-m_Q\right)^2 = \pi\sigma L. \label{eq:CF3}
\end{eqnarray}
The same form of this equation was derived in Ref.\cite{Veseli:1996gy} by using the simplified potential model with a couple of intuitive approximations.

Modified and elaborated forms of Eq. (\ref{eq:CF3}) have been proposed in Refs. \cite{Selem:2006nd, Chen:2014nyo}, among which Ref.~\cite{Selem:2006nd} gives
\begin{eqnarray}
  M=m_Q+\sqrt{\frac{\sigma' L}{2}} + 2^{1/4}\kappa L^{-1/4}m_q^{3/2},
  \label{wilczek}
\end{eqnarray}
where $\sigma'=2\pi\sigma$ and $\kappa=$ a parameter determined by a computer simulation. Reference~\cite{Chen:2014nyo} modifies Eq.~(\ref{wilczek}) so that it does not have singularity at $L=0$ and gives,
\begin{eqnarray}
  \left(M-m_Q\right)^2= \frac{\sigma' L}{2} + \left(m_q+m_Qv_2^2\right)^2,
  \label{bing}
\end{eqnarray}
where $v_2$ is a velocity of the heavy quark in a heavy-light hadron system.
From our point of view in this paper, the last terms of Eqs.~(\ref{wilczek}) and (\ref{bing}) are not necessary to analyze the experimental data of heavy-light systems in order to compare them with theoretical models since data or theoretical values with the same $L$ are averaged over isospin and angular momentum $L$. {However, we allow a constant term on the right hand side of Eqs.~(\ref{wilczek}) or (\ref{bing}) as later shown in Eq. (\ref{eq:coefDB}).} Reference \cite{Chen:2014nyo} further studied detailed mass spectra of $\Lambda_Q$ and $\Xi_Q$ by including $LS$ coupling.

\section{Numerical plots for heavy-light systems and universal description}\label{sec3}

According to Eq. (\ref{eq:CF3}), we plot figures for heavy-light mesons, $D/B/D_s/B_s$, as well as charmed and bottom baryons, $\Lambda_Q/\Sigma_Q/\Xi_Q/\Xi'_Q/\Omega_Q$ with $Q=c, b$, for experimental data listed in PDG \cite{Agashe:2014kda} and some theoretical models Refs. \cite{Ebert:2009ua,Ebert:2011kk,Chen:2017gnu,Shah:2016nxi,Shah:2016mig}. In the following, when plotting experimental data, we adopt the quark masses used in Ref. \cite{Ebert:2009ua,Ebert:2011kk} in Eq.~(\ref{eq:CF3}) as
\begin{eqnarray}
  m_c=1.55 {\rm ~GeV},\quad m_b = 4.88 {\rm ~GeV}.
  \label{eq:massq}
\end{eqnarray}
This is because all the figures with expeimental data are compared with model calculation given by Ref. \cite{Ebert:2009ua,Ebert:2011kk}.
In the case of $\Omega_Q$ baryons, we plot other model calculations \cite{Chen:2017gnu,Shah:2016nxi,Shah:2016mig}.

\subsection{Heavy-Light Mesons}\label{sec3-1}

In this subsection, we plot figures in $(M-m_Q)^2$ versus $L$ for $D$, $B$, $D_s$, and $B_s$ mesons taken from experimental data \cite{Agashe:2014kda} as well as model calculations of Ref. \cite{Ebert:2009ua}.
\begin{table}[htbp]
\caption{The experimental \cite{Agashe:2014kda} and EFG model \cite{Ebert:2009ua} values for $D$ and $B$ to obtain Figs. \ref{expfigDB}, \ref{modelfigDB} and \ref{DBmesons}.
}
\label{MGIFF1}
\centering
\begin{tabular}{c|ccc|cccccccc}
\toprule[1pt]
\toprule[1pt]
State&\multicolumn{2}{c}{Experiment}&EFG&\multicolumn{2}{c}{Experiment}&EFG\\
\midrule
$|n^{2S+1}L_J,J^{P}\rangle$&meson&mass&$c\bar{q}$&meson&mass&$b\bar{q}$\\
\midrule[1pt]
$|1{}^{1}S_0,0^-\rangle$&$D$&$1869.59$&1871&$B$&5279.6&5280\\
$|1{}^{3}S_1,1^{-}\rangle$&$D^*(2010)$&$2010.26$&2010&$B^*$&5324.7&5326\\
$|1{}^{3}P_0,0^+\rangle$&$D_0^*(2400)$&$2351$&2406&$B_J^*(5732)$&5698&5749\\
$|1P_1,1^+\rangle$&$D_1(2430)$&$2427$&$2469$&&&5774\\
$|1P_1,1^+\rangle$&$D_1(2420)$&$2423.2$&2426&$B_1(5721)$&5726.0&5723\\
$|1{}^3P_2,2^+\rangle$&$D_2^*(2460)$&$2465.4$&2460&$B_2^*(5747)$&5739.5&5741\\
$|1{}^3D_1,1^-\rangle$&&&2788&&&6119\\
$|1D_2,2^-\rangle$&&&2850&&&6121\\
$|1D_2,2^-\rangle$&&&2806&&&6103\\
$|1{}^3D_3,3^-\rangle$&&&$2863$&&&6091\\
$|1{}^3F_2,2^+\rangle$&&&3090&&&6412\\
$|1F_3,3^+\rangle$&&&3145&&&6420\\
$|1F_3,3^+\rangle$&&&3129&&&6391\\
$|1{}^3F_4,4^+\rangle$&&&3187&&&6380\\
$|1^{3}G_3,3^-\rangle$&&&3352&&&6664\\
$|1G_4,4^-\rangle$&&&3415&&&6652\\
$|1G_4,4^-\rangle$&&&3403&&&6648\\
$|1{}^3G_5,5^-\rangle$&&&$3473$&&&6634\\
\bottomrule[1pt]
\end{tabular}
\end{table}

\begin{table}[htbp]
\caption{The experimental \cite{Agashe:2014kda} and EFG model \cite{Ebert:2009ua} values for $D_s$ and $B_s$ to obtain Figs. \ref{expfigDB}, \ref{modelfigDB} and \ref{DBmesons}.
}
\label{MGIFF2}
\centering
\begin{tabular}{c|ccc|cccccccc}
\toprule[1pt]
\toprule[1pt]
State&\multicolumn{2}{c}{Experiment}&EFG&\multicolumn{2}{c}{Experiment}&EFG\\
\midrule
$|n^{2S+1}L_J,J^{P}\rangle$&meson&mass&$c\bar{s}$&meson&mass&$b\bar{s}$\\
\midrule[1pt]
$|1{}^{1}S_0,0^-\rangle$&$D_s$&$1969.0$&1969&$B_s$&5366.9&5372\\
$|1{}^{3}S_1,1^{-}\rangle$&$D_s^*$&$2112.1$&2111&$B_s^*$&5415.4&5414\\
$|1{}^{3}P_0,0^+\rangle$&$D_{s0}^*(2317)$&$2318.0$&$2509$&&&5833\\
$|1P_1,1^+\rangle$&$D_{s1}(2460)$&$2459.6$&2574&$B_{sJ}^*(5850)$&5853&5865\\
$|1P_1,1^+\rangle$&$D_{s1}(2536)$&$2535.18$&2536&$B_{s1}(5830)$&5828.6&5831\\
$|1{}^3P_2,2^+\rangle$&$D_{s2}(2573)$&$2569.1$&2571&$B^*_{s2}(5840)$&5839.9&5842\\
$|1{}^3D_1,1^-\rangle$&&&2913&&&6209\\
$|1D_2,2^-\rangle$&&&2961&&&6218\\
$|1D_2,2^-\rangle$&&&2931&&&6189\\
$|1{}^3D_3,3^-\rangle$&$D_{sJ}^*(2860)$&$2860.5$&$2971$&&&6191\\
$|1{}^3F_2,2^+\rangle$&&&3230&&&6501\\
$|1F_3,3^+\rangle$&&&3266&&&6515\\
$|1F_3,3^+\rangle$&&&3254&&&6468\\
$|1{}^3F_4,4^+\rangle$&&&3300&&&6475\\
$|1^{3}G_3,3^-\rangle$&&&3508&&&6753\\
$|1G_4,4^-\rangle$&&&3554&&&6762\\
$|1G_4,4^-\rangle$&&&3546&&&6715\\
$|1{}^3G_5,5^-\rangle$&&&$3595$&&&6726\\
\bottomrule[1pt]
\end{tabular}
\end{table}

\begin{figure*}[htpb]
	\begin{center}
		\includegraphics[scale=0.6]{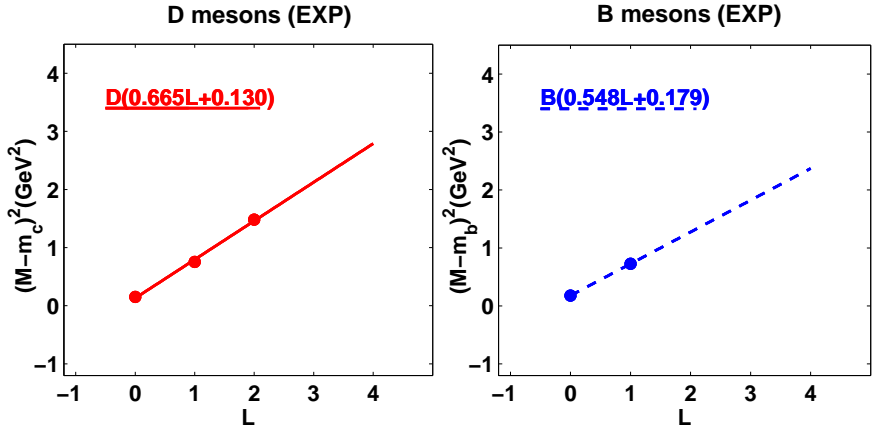}
		\caption{(color online). Plots of experimental data for $D$ and $B$ mesons in $L$ vs. $(M-m_{c,b})^2$. The best fit lines are given with equations.}\label{expfigDB}
	\end{center}
\end{figure*}

\begin{figure*}[htpb]
	\begin{center}
		\includegraphics[scale=0.675]{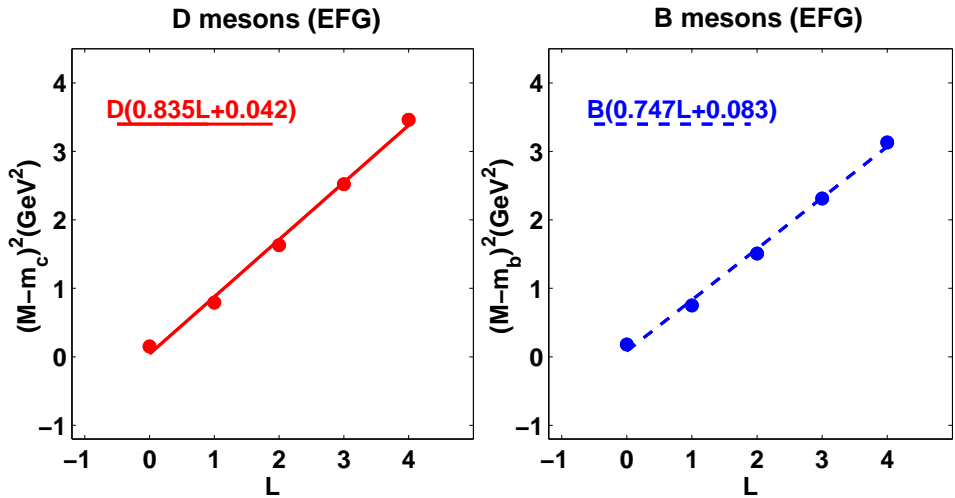}
		\caption{(color online). Plots of values calculated in Ref. \cite{Ebert:2009ua} similar to Fig. \ref{expfigDB}. The best fit lines are given with equations.}\label{modelfigDB}
	\end{center}
\end{figure*}

$D/B$ mesons :  Using Tables \ref{MGIFF1} and \ref{MGIFF2}, the results are given in Figs. \ref{expfigDB} and \ref{modelfigDB}for $D$ and $B$ mesons, separately.
To compare Eq. (\ref{eq:CF1}) with Eq. (\ref{eq:CF3}), we give a numerical value of the coefficient of $L$ in Eq. (\ref{eq:CF1}) obtained by Afonin,
\begin{eqnarray}
  M^2 = 1.103L + 1.102n+0.686.
  \label{slopeLight}
\end{eqnarray}
which is taken from Table 4 of Ref. \cite{Afonin:2007jd} with principal quantum number $n$. The calculations of the EFG model \cite{Ebert:2009ua} for $D$ and $B$ mesons are given by Fig. \ref{modelfigDB}.

\begin{figure*}[htpb]
	\begin{center}
		\includegraphics[scale=0.65]{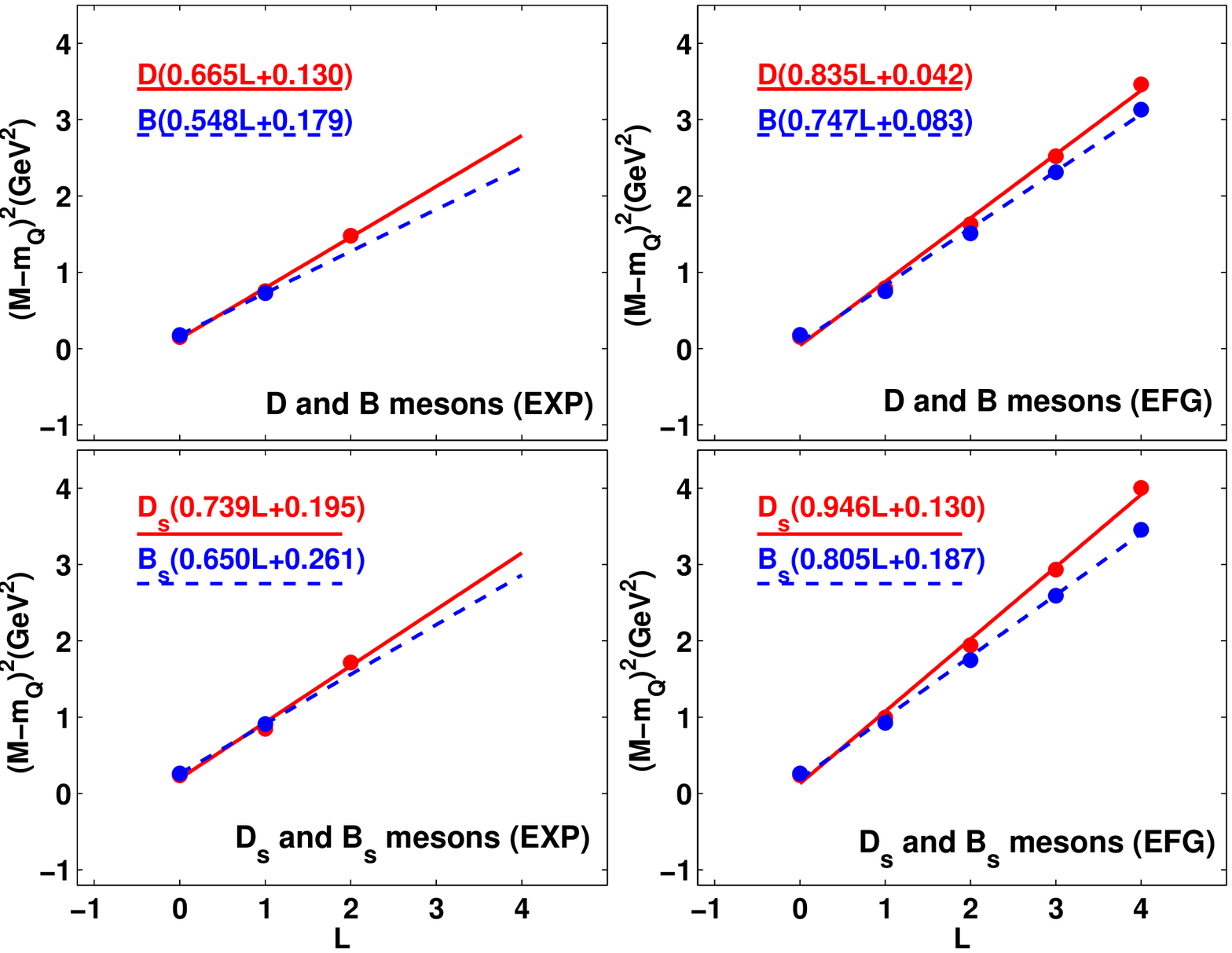}
	\caption{(color online). Plots of experimental data and EFG model \cite{Ebert:2009ua} calculation both for $D_{(s)}$ and $B_{(s)}$ mesons. The best fit lines are given with equations.}\label{DBmesons}
	\end{center}
\end{figure*}

Looking at linear equations written on Fig. \ref{expfigDB}, we obtain
\begin{equation}
\begin{split}
  \left(M_D-m_c\right)^2 &= 0.665L + 0.130, \\
  \left(M_B-m_b\right)^2 &= 0.548L + 0.179. \label{eq:coefDB}
\end{split}
\end{equation}
As you can see, coefficients of $L$ are nearly equal to 1/2 of that of light hadrons in Eq. (\ref{slopeLight}).
Hence we can conclude that $D$ and $B$ mesons satisfy Eq. (\ref{eq:CF3}), support an approximate rotational symmetry of heavy-light mesons claimed in the previous paper \cite{Matsuki:2016hzk}, and the string picture for heavy-light mesons well works. Two lines in Fig. \ref{expfigDB} can be written in one figure in the upper row of Fig.~\ref{DBmesons}, which nicely shows a universal description of these mesons. This means that two lines almost overlap with each other irrespective of heavy quark flavors. In the same way, we draw a figure for the EFG model \cite{Ebert:2009ua} which is also given in the upper right of Fig. \ref{DBmesons}. From hereon, we plot two lines for heavy-light systems $X_c$ and $X_b$ in one figure, which makes easy to compare both lines and we can extract conclusions on whether their slopes are close to 1/2 and whether they overlap to confirm a universal description.
Finally, we predict the average mass of $B$ with $L=2$ as 6.009 GeV using Eq.~(\ref{eq:coefDB}).

$D_s/B_s$ mesons : In the lower row of Fig.~\ref{DBmesons}, we plot mass squared versus $L$ for $D_s$ and $B_s$ mesons using Tables \ref{MGIFF1} and \ref{MGIFF2}, which presents similar behavior to that of $D$ and $B$ mesons, it is obvious that they also satisfy Eq.~(\ref{eq:CF3}). The values of slope for $D_s$ and $B_s$ are close to each other, which means $D_s$ and $B_s$ satisfy a universal description.
To show model calculations for $D_s/B_s$ mesons, we plot the figure for these mesons of the GI model in the lower right of Fig.~\ref{DBmesons}, which shows the GI model well satisfies a linear equation Eq.~(\ref{eq:CF3}) of $(M-m_Q)^2$ versus $L$ and two lines almost overlap, i.e., a universal description is confirmed. As one can see, the experimental data is much better than the model calculation in regard to a universal description, which may be due to the fact that the EFG model does not explicitly respect heavy quark symmetry.
We also predict the average mass of $B_s$ with $L=2$ as 6.129 GeV using $\left(M_{B_s}-m_b\right)^2 = 0.650L + 0.261$ written on the lower left of Fig. \ref{DBmesons}.
%

%

\subsection{Charmed and Bottom Baryons}\label{sec3-2}

Regarding two light quarks inside a heavy-baryon as a light cluster (diquark), we can apply the formula Eq.~(\ref{eq:CF3}) to charmed and bottom baryons, in which only one of $c$ or $b$ quark is included. For the baryons, we take models of Refs. \cite{Ebert:2011kk,Chen:2017gnu,Shah:2016mig,Shah:2016nxi} to compare with experiments \cite{Agashe:2014kda}, which has including the effect of heavy quark symmetry. Here we should mention that there are other models for calculating mass spectrum of heavy baryons as well as a pioneering work \cite{{Capstick:1986bm,Pervin:2007wa,Roberts:2007ni}}. We also have to take care of a total spin of a diquark, $S_{qq}$. We call this system a {\it good type} when $S_{qq}=0$, and a {\it bad type} when $S_{qq}=1$ according to Ref. \cite{Jaffe:2004ph}.

\begin{figure*}[htpb]
	\begin{center}
		\includegraphics[scale=0.65]{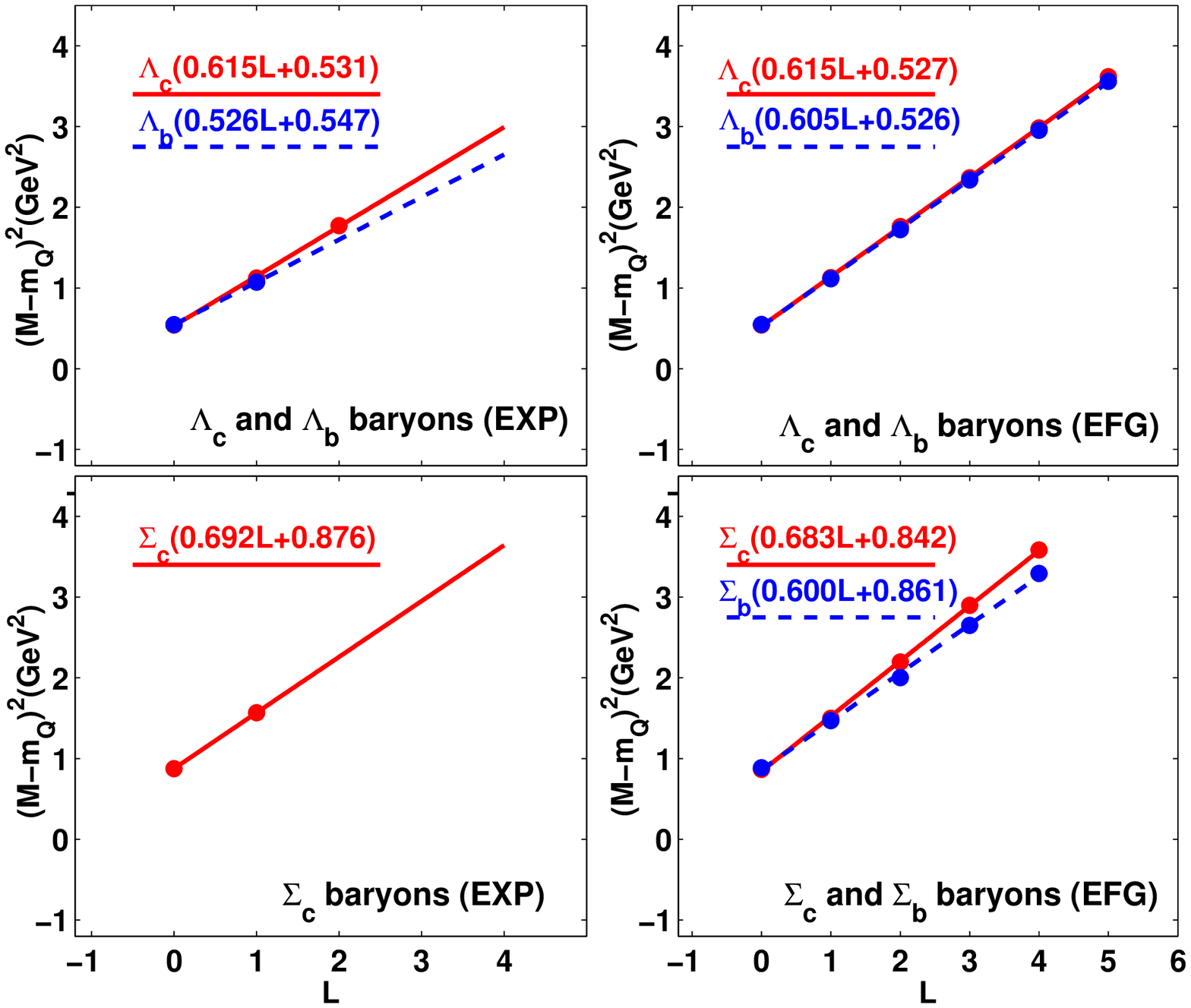}
		\caption{(color online). Plots of experimental data and model calculation \cite{Ebert:2011kk} for $\Lambda_Q$ and $\Sigma_Q$ baryons. The best fit lines are given with equations.}\label{lambdasigmas}
	\end{center}
\end{figure*}

$\Lambda_c/\Lambda_b$ baryons ($I=0$, $S_{qq}=0$) : $\Lambda_Q$ baryons have $S_{qq}=0$, i.e., a good type. Charmed and bottom $\Lambda_Q$ baryons have isospin $I=0$ and $S_{qq}=0$. In Table \ref{gooddiquark}, we listed the present experimental data and EFG model calculated results for $\Lambda_Q$ baryons. Their Regge-like lines are given by the upper row of Fig.~\ref{lambdasigmas} for experimental and theoretical values. We take theoretical values from Ref. \cite{Ebert:2011kk}. From these figures, we can conclude that Eq.~(\ref{eq:CF3}) is satisfied, the slopes are close to 1/2, and unified description holds. We can predict the average values of some states which are not yet observed. For instance, by using the line for $\Lambda_b$, we predict the average mass of $\Lambda_b(3/2^+,5/2^+)$ with $L=2$ as 6.145 GeV.

\begin{table*}[htbp]
\caption{The experimental \cite{Agashe:2014kda} and EFG model values \cite{Ebert:2011kk} for $\Lambda_Q$ and $\Xi_Q$ baryons with a good diquark ($S_{qq}=0$). Here, the constituent quark masses \cite{Ebert:2009ua} we adopted to analyze experimental data as well as EFG calculations are given by Eq.~(\ref{eq:massq}).}
\label{gooddiquark}
\centering
\begin{tabular}{c|ccc|ccc|c|ccc|ccccccccc}
\toprule[1pt]
&\multicolumn{3}{|c|}{$\Lambda_c$}&\multicolumn{3}{|c|}{$\Lambda_b$}&&\multicolumn{3}{|c|}{$\Xi_c$}&\multicolumn{3}{c}{$\Xi_b$}\\
\midrule[1pt]
State&\multicolumn{2}{c}{Experiment}&EFG&\multicolumn{2}{c}{Experiment}&EFG&State&\multicolumn{2}{c}{Experiment}&EFG&\multicolumn{2}{c}{Experiment}&EFG\\
$|nL,I(J^P)\rangle$&bayron&$M^{exp}$&$M$&bayron&$M^{exp}$&$M$&$|nL,I(J^P)\rangle$&bayron&$M^{exp}$&$M$&bayron&$M^{exp}$&$M$\\
\midrule[1pt]
$|1S,0(\frac{1}{2}^+)\rangle$&$\Lambda_c(2286)$&2286.46&2286&$\Lambda_b(5620)$&5619.58&5620&$|1S,\frac{1}{2}(\frac{1}{2}^+)\rangle$&$\Xi_c(2470)$&2470.87&2476&$\Xi_b(5790)$&5794.5&5803\\
$|1P,0(\frac{1}{2}^-)\rangle$&$\Lambda_{c}(2595)$&$2592.25$&2598&&&5930&$|1P,\frac{1}{2}(\frac{1}{2}^-)\rangle$&$\Xi_c(2790)$&2792.0&2792&&&6120\\
$|1P,0(\frac{3}{2}^-)\rangle$&$\Lambda_c(2628)$&2628.11&2627&&&5942&$|1P,\frac{1}{2}(\frac{3}{2}^-)\rangle$&$\Xi_c(2815)$&$2816.67$&2819&&&6130\\
$|1D,0(\frac{3}{2}^+)\rangle$&&&2874&&&6190&$|1D,\frac{1}{2}(\frac{3}{2}^+)\rangle$&$\Xi_c(3055)$&3055.9&3059&&&6366\\
$|1D,0(\frac{5}{2}^+)\rangle$&$\Lambda_c(2880)$&2881.53&2880&&&6196&$|1D,\frac{1}{2}(\frac{5}{2}^+)\rangle$&$\Xi_c(3080)$&3077.2&3076&&&6373\\
$|1F,0(\frac{5}{2}^-)\rangle$&&&3097&&&6408&$|1F,\frac{1}{2}(\frac{5}{2}^-)\rangle$&&&3278&&&6577\\
$|1F,0(\frac{7}{2}^-)\rangle$&&&3078&&&6411&$|1F,\frac{1}{2}(\frac{7}{2}^-)\rangle$&&&$3292$&&&6581\\
$|1G,0(\frac{7}{2}^+)\rangle$&&&3270&&&6598&$|1G,\frac{1}{2}(\frac{7}{2}^+)\rangle$&&&3469&&&6760\\
$|1G,0(\frac{9}{2}^+)\rangle$&&&3284&&&6599&$|1G,\frac{1}{2}(\frac{9}{2}^+)\rangle$&&&3483&&&6762\\
$|1H,0(\frac{9}{2}^-)\rangle$&&&3444&&&6767&$|1H,\frac{1}{2}(\frac{9}{2}^-)\rangle$&&&3643&&&6933\\
$|1H,0(\frac{11}{2}^-)\rangle$&&&3460&&&6766&$|1H,\frac{1}{2}(\frac{11}{2}^-)\rangle$&&&3658&&&6934\\
\bottomrule[1pt]
\end{tabular}
\end{table*}

$\Sigma_c/\Sigma_b$ baryons ($I=1$, $S_{qq}=1$) :
$\Sigma_Q$ baryons have $S_{qq}=1$, i.e., a bad type. We present the experimental data \cite{Agashe:2014kda} and EFG model results \cite{Ebert:2011kk} for $\Sigma_Q$ baryons in Table \ref{baddiquark}. From this Table, we can see that only two experimental data for $\Sigma_b$ with $L=0$ were observed, i.e., only one point in the $L$ vs. $(M-m_b)^2$ plane, and experimental data for $\Sigma_c$ with $L=0, 1$ have been measured. Hence, we plot figures only for $\Sigma_c$ of experimental and $\Sigma_{c,b}$ of theoretical values \cite{Ebert:2011kk} in the lower row of Fig. \ref{lambdasigmas}. From these figures, we can conclude again that Eq.~(\ref{eq:CF3}) is satisfied for $\Sigma_c$, the slope is close to 1/2, unified description holds, and the model calculations obey the same rules as experimental data, too.
We predict the average mass of $\Sigma_c$ with $L=2$ as 3.053 GeV using the linear equation for $\Sigma_c$ written on the lower left of Fig. \ref{lambdasigmas}.

\begin{table*}[htbp]
\caption{The experimental \cite{Agashe:2014kda} and EFG model values \cite{Ebert:2011kk} for $\Sigma_Q$ and $\Xi^{\prime}_Q$ baryons with a bad diquark ($S_{qq}=1$). Here, the constituent quark masses \cite{Ebert:2009ua} we adopted to analyze experimental data as well as EFG calculations are given by Eq.~(\ref{eq:massq}).}
\label{baddiquark}
\centering
\begin{tabular}{c|ccc|ccc|c|ccc|ccccccccc}
\toprule[1pt]
&\multicolumn{3}{c|}{$\Sigma_c$}&\multicolumn{3}{c|}{$\Sigma_b$}&&\multicolumn{3}{c|}{$\Xi^{\prime}_c$}&\multicolumn{3}{c}{$\Xi^{\prime}_b$}\\
\midrule[1pt]
State&\multicolumn{2}{c}{Experiment}&EFG&\multicolumn{2}{c}{Experiment}&EFG&State&\multicolumn{2}{c}{Experiment}&EFG&\multicolumn{2}{c}{Experiment}&EFG\\
$|nL,I(J^P)\rangle$&bayron&$M^{exp}$&$M$&bayron&$M^{exp}$&$M$&$|nL,I(J^P)\rangle$&bayron&$M^{exp}$&$M$&bayron&$M^{exp}$&$M$\\
\midrule[1pt]
$|1S,1(\frac{1}{2}^+)\rangle$&$\Sigma_c(2455)$&2453.75&2443&$\Sigma_b(5810)$&5811.3&5808&$|1S,\frac{1}{2}(\frac{1}{2}^+)\rangle$&$\Xi_c(2579)$&2577.4&2579&$\Xi_b(5935)$&5935.02&5936\\
$|1S,1(\frac{3}{2}^+)\rangle$&$\Sigma_c(2520)$&2518.48&$2519$&$\Sigma_b(5830)$&$5832.1$&5834&$|1S,\frac{1}{2}(\frac{1}{2}^+)\rangle$&$\Xi_c(2645)$&2645.53&2649&&&5963\\
$|1P,1(\frac{1}{2}^-)\rangle$&$\Sigma_c(2800)$&2806&2799&&&6101&$|1P,\frac{1}{2}(\frac{1}{2}^-)\rangle$&$\Xi_c(2930)$&2931&2936&&&6233\\
$|1P,1(\frac{1}{2}^-)\rangle$&&&2713&&&6095&$|1P,\frac{1}{2}(\frac{1}{2}^-)\rangle$&&&2854&&&6227\\
$|1P,1(\frac{3}{2}^-)\rangle$&$\Sigma_c(2800)$&2806&2798&&&6096&$|1P,\frac{1}{2}(\frac{3}{2}^-)\rangle$&$\Xi_{c}(2930)$&2931&2935&&&6234\\
$|1P,1(\frac{3}{2}^-)\rangle$&&&2773&&&6087&$|1P,\frac{1}{2}(\frac{3}{2}^-)\rangle$&&&2912&&&6224\\
$|1P,1(\frac{5}{2}^-)\rangle$&&&2789&&&6084&$|1P,\frac{1}{2}(\frac{5}{2}^-)\rangle$&$\Xi_{c}(2930)$&2931&2929&&&6226\\
$|1D,1(\frac{1}{2}^+)\rangle$&&&3041&&&6311&$|1D,\frac{1}{2}(\frac{1}{2}^+)\rangle$&&&3163&&&6447\\
$|1D,1(\frac{3}{2}^+)\rangle$&&&3043&&&6326&$|1D,\frac{1}{2}(\frac{3}{2}^+)\rangle$&&&3167&&&6459\\
$|1D,1(\frac{3}{2}^+)\rangle$&&&3040&&&6285&$|1D,\frac{1}{2}(\frac{3}{2}^+)\rangle$&&&3160&&&6431\\
$|1D,1(\frac{5}{2}^+)\rangle$&&&3038&&&6284&$|1D,\frac{1}{2}(\frac{3}{2}^+)\rangle$&&&3166&&&6432\\
$|1D,1(\frac{5}{2}^+)\rangle$&&&3023&&&6270&$|1D,\frac{1}{2}(\frac{5}{2}^+)\rangle$&&&3153&&&6420\\
$|1D,1(\frac{7}{2}^+)\rangle$&&&3013&&&6260&$|1D,\frac{1}{2}(\frac{7}{2}^+)\rangle$&$\Xi_c(3123)$&3122.9&3147&&&6414\\
$|1F,1(\frac{3}{2}^-)\rangle$&&&3288&&&6550&$|1F,\frac{1}{2}(\frac{3}{2}^-)\rangle$&&&3418&&&6675\\
$|1F,1(\frac{5}{2}^-)\rangle$&&&3283&&&6564&$|1F,\frac{1}{2}(\frac{5}{2}^-)\rangle$&&&3408&&&6686\\
$|1F,1(\frac{5}{2}^-)\rangle$&&&3254&&&6501&$|1F,\frac{1}{2}(\frac{5}{2}^-)\rangle$&&&3394&&&6640\\
$|1F,1(\frac{7}{2}^-)\rangle$&&&3253&&&6500&$|1F,\frac{1}{2}(\frac{7}{2}^-)\rangle$&&&3393&&&6641\\
$|1F,1(\frac{7}{2}^-)\rangle$&&&3227&&&6472&$|1F,\frac{1}{2}(\frac{7}{2}^-)\rangle$&&&3373&&&6619\\
$|1F,1(\frac{9}{2}^-)\rangle$&&&3209&&&6459&$|1F,\frac{1}{2}(\frac{9}{2}^-)\rangle$&&&3357&&&6610\\
$|1G,1(\frac{5}{2}^+)\rangle$&&&3495&&&6749&$|1G,\frac{1}{2}(\frac{5}{2}^+)\rangle$&&&3623&&&6867\\
$|1G,1(\frac{7}{2}^+)\rangle$&&&3483&&&6761&$|1G,\frac{1}{2}(\frac{7}{2}^+)\rangle$&&&3608&&&6876\\
$|1G,1(\frac{7}{2}^+)\rangle$&&&3444&&&6688&$|1G,\frac{1}{2}(\frac{7}{2}^+)\rangle$&&&3584&&&6822\\
$|1G,1(\frac{9}{2}^+)\rangle$&&&3442&&&6687&$|1G,\frac{1}{2}(\frac{9}{2}^+)\rangle$&&&3582&&&6821\\
$|1G,1(\frac{9}{2}^+)\rangle$&&&3410&&&6648&$|1G,\frac{1}{2}(\frac{9}{2}^+)\rangle$&&&3558&&&6792\\
$|1G,1(\frac{11}{2}^+)\rangle$&&&3386&&&6635&$|1G,\frac{1}{2}(\frac{11}{2}^+)\rangle$&&&3536&&&6782\\
\bottomrule[1pt]
\end{tabular}
\end{table*}

\begin{figure*}[htpb]
	\begin{center}
		\includegraphics[scale=0.6]{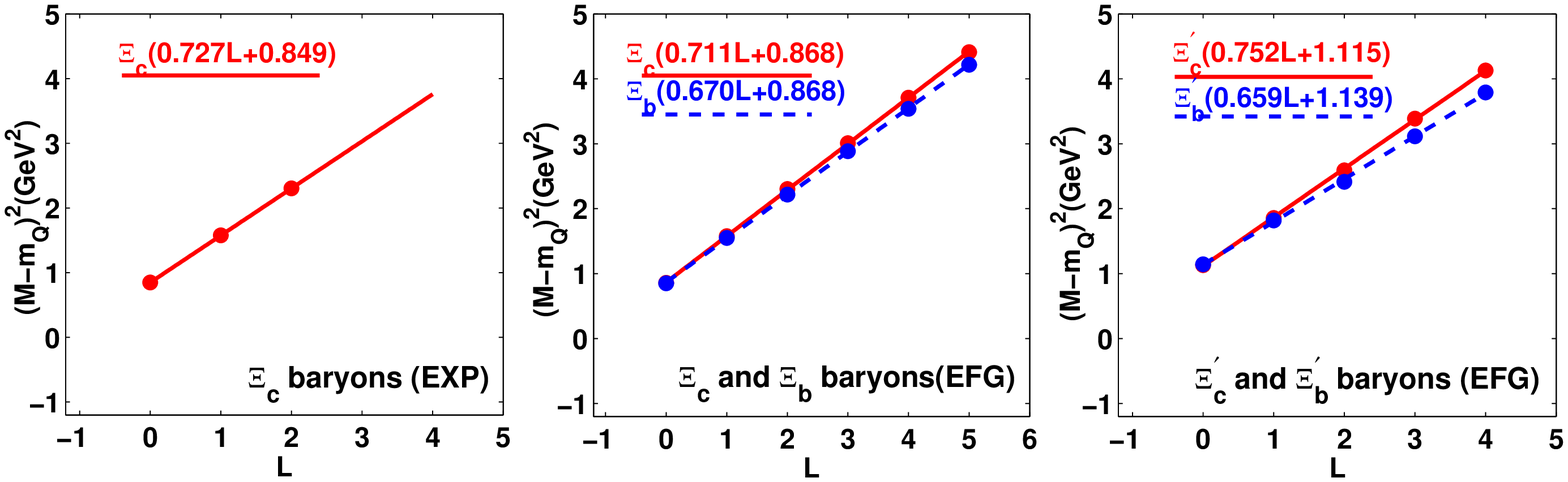}
		\caption{(color online). Plots of experimental data for $\Xi_c$ baryons and model calculation results \cite{Ebert:2011kk} of $\Xi_{Q}$ and $\Xi^{\prime}_{Q}$ baryons. The best fit lines are given with equations.}\label{Xics}
	\end{center}
\end{figure*}

$\Xi_c/\Xi_b$ baryons ($I=1/2$, $S_{qq}=0$) :
$\Xi_Q$ baryons have $S_{qq}=0$, i.e., a good type. The experimental data and theoretical values for $\Xi_Q$ baryons are listed in Table \ref{gooddiquark}.
Because of lack of experimental data, we can only plot a figure for experimental values of $\Xi_c$ with $L=0, 1, 2$.
Theoretical values of Ref. \cite{Ebert:2011kk} are taken. Their figures are given by the left two of Fig. \ref{Xics}, respectively. From these figures, we can conclude again that Eq.~(\ref{eq:CF3}) is satisfied for $\Xi_c$ and the slope is close to 1/2.
The model calculations obey the same rules as experimental data, too, including unified description.
We predict the average mass of $\Xi_c$ with $L=2$ as 3.068 GeV using the linear equation for $\Xi_c$ written on the leftmost of Fig. \ref{Xics}.

$\Xi'_c/\Xi'_b$ baryons ($I=1/2$, $S_{qq}=1$) :
$\Xi'_Q$ baryons have $S_{qq}=1$, i.e., a bad type. We also present the experimental data and model calculation for $\Xi^{\prime}_Q$ baryons in Table \ref{baddiquark}.
There is no definite experimental data for $\Xi'_Q$ with $Q=c, b$, and hence, we can only plot figures for $\Xi'_Q$ of model calculations given by Ref. \cite{Ebert:2011kk}. The figure is given in the rightmost one of Fig.~\ref{Xics}, which shows that Eq.~(\ref{eq:CF3}) is satisfied, the slope is close to 1/2, and unified description holds for theoretical data.

\begin{figure*}[htpb]
	\begin{center}
		\includegraphics[scale=0.65]{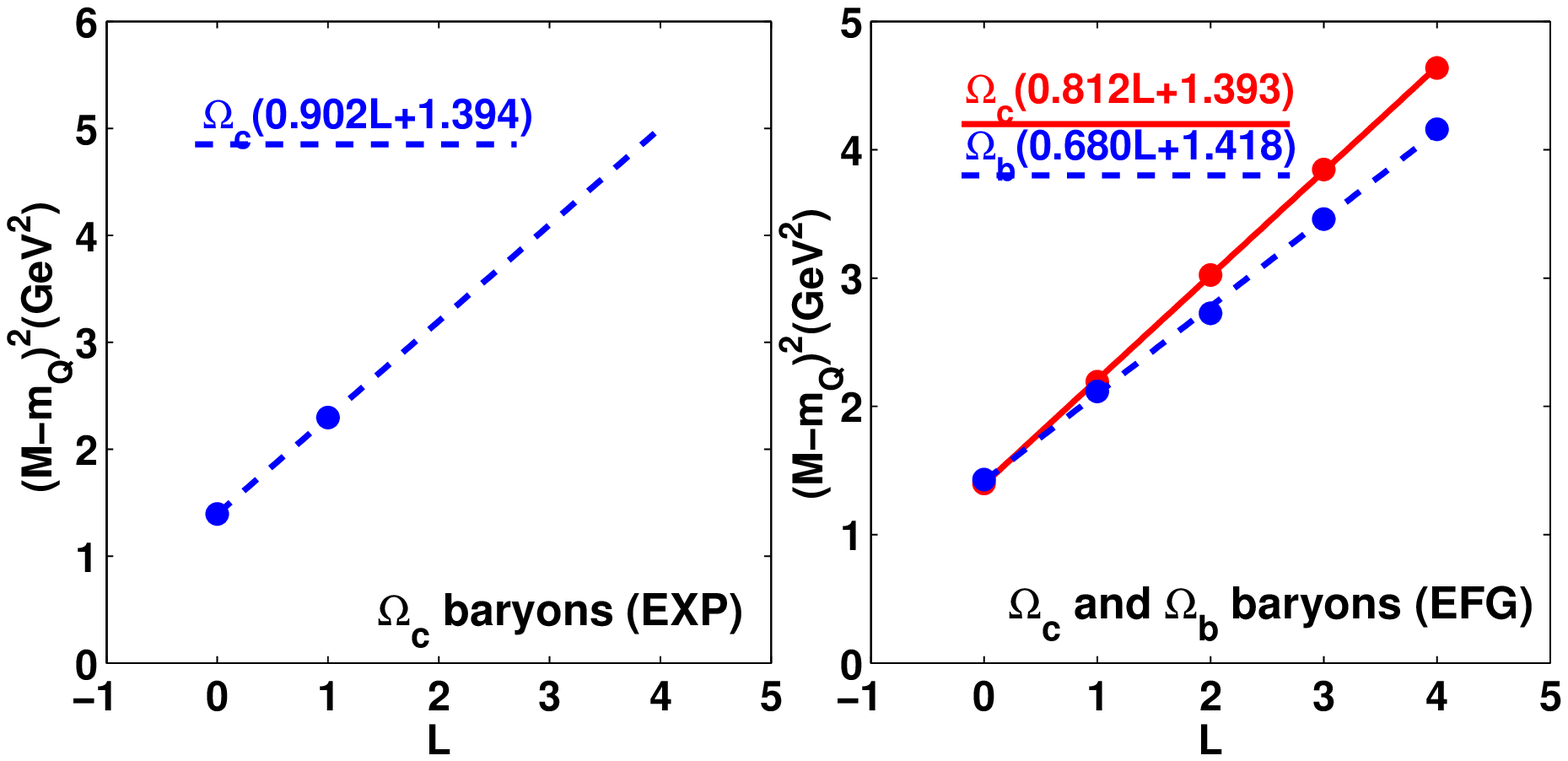}
		\caption{(color online). Plots of experimental data for $\Omega_c$ baryons and model calculation results \cite{Ebert:2011kk} of $\Omega_{Q}$ baryons. The best fit lines are given with equations.}\label{omegas}
	\end{center}
\end{figure*}

\begin{figure*}[htpb]
	\begin{center}
		\includegraphics[scale=0.65]{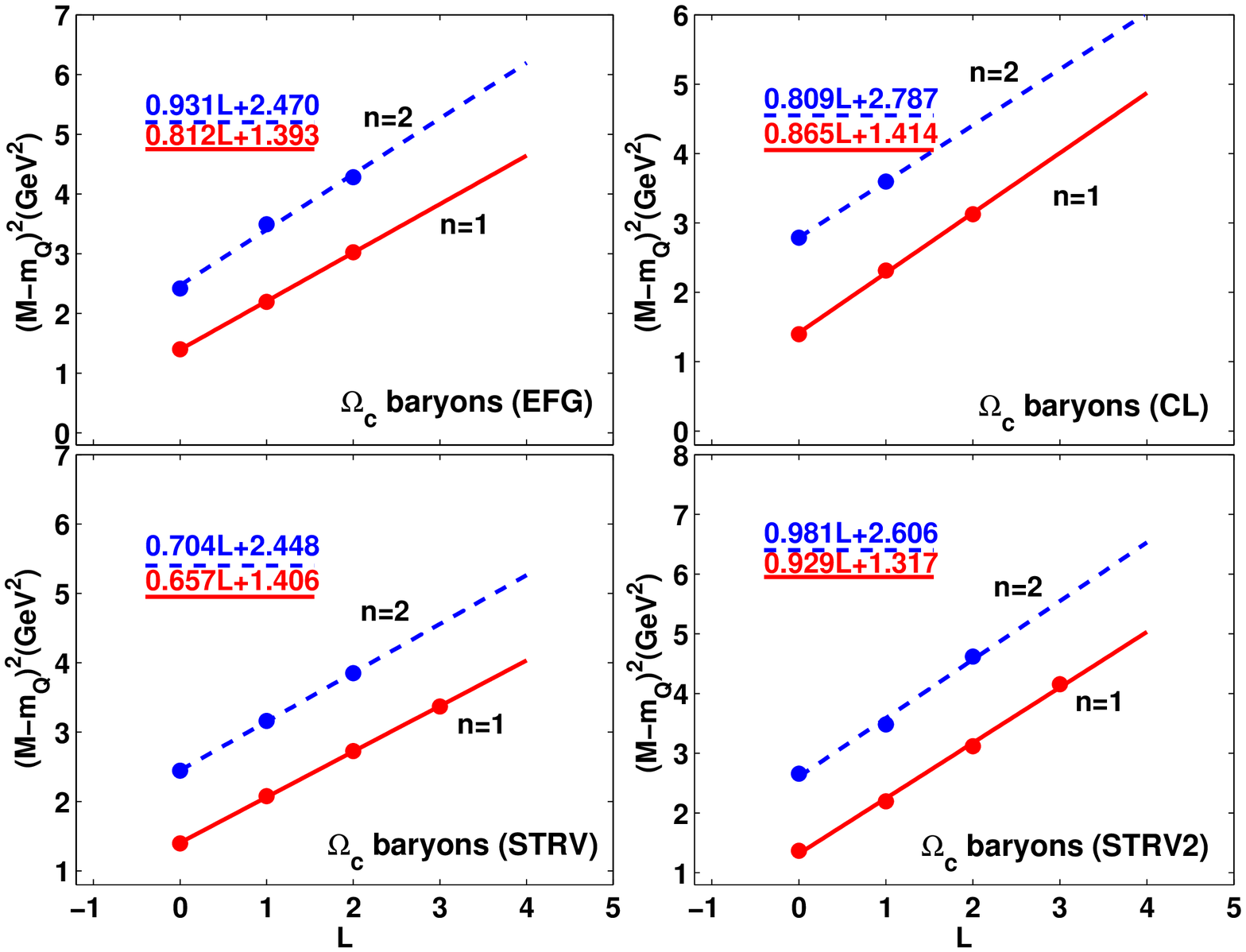}
		\caption{(color online). Plots of  values calculated in $EFG$ \cite{Ebert:2011kk}, $CL$ \cite{Chen:2017gnu}, $STRV$ \cite{Shah:2016nxi}, $STRV2$ \cite{Shah:2016mig} models for $\Omega_Q$ baryons with principal quantum number $n=1, 2$. The best fit lines are given with an equation.}\label{omegacn12s}
	\end{center}
\end{figure*}

$\Omega_c/\Omega_b$ baryons ($I=0$, $S_{qq}=1$) :
$\Omega_Q$ baryons have $S_{qq}=1$, i.e., a bad type. Recently the LHCb Collaboration ovserved six $\Omega_c^0$ in the $\Xi_c^+K^-$ invariant mass spectrum \cite{Aaij:2017nav}. However, there is only one $\Omega_b$ state with $L=0$ listed in \cite{Agashe:2014kda}. Hence, we only list the present experimental data and some of the theoretical results in Table \ref{Omegadata}. We also plot the left of Fig. \ref{omegas} for $\Omega_c$ of experimental data and the right of Fig. \ref{omegas} for $\Omega_c$ and $\Omega_b$ of the theoretical calculation in Ref. \cite{Ebert:2011kk}. Besides, since there might be higher radial excited states in the recent discovery by LHCb, we also plot figures with principal quantum number $n=1, 2$ for theoretical values \cite{Ebert:2011kk,Shah:2016nxi,Shah:2016mig,Chen:2017gnu} in Fig. \ref{omegacn12s}.
We predict the average mass of $\Omega_c$ with $L=2$ as 3.338 GeV using the linear equation for $\Omega_c$ written on the left of Fig. \ref{omegas}.

\begin{table}[htbp]
\caption{The experimental and different model calculated values for $\Omega_c$ baryons with a bad diquark ($S_{qq}=1$). Here, the constituent quark masses \cite{Ebert:2009ua} we adopted to analyze experimental data as well as EFG calculations are given by Eq.~(\ref{eq:massq}).}
\label{Omegadata}
\centering
\begin{tabular}{cccccccccccccccccccc}
\toprule[1pt]
\toprule
State&EXP \cite{Agashe:2014kda,Aaij:2017nav}&CL \cite{Chen:2017gnu} &EFG \cite{Ebert:2011kk}&STRV \cite{Shah:2016nxi}&STRV2 \cite{Shah:2016mig}\\
\midrule[1pt]
$|1S,\frac{1}{2}^+\rangle$&2695.2&2698&2698&2695&2695\\
$|1S,\frac{3}{2}^+\rangle$&2765.9&2765&2768&2767&2745\\
$|1P,\frac{1}{2}^-\rangle$&3000.4&3033&2966&3011&3041\\
$|1P,\frac{1}{2}^-\rangle$&3065.6&3075&3055&3028&3050\\
$|1P,\frac{3}{2}^-\rangle$&3050.2&3068&3029&2976&3024\\
$|1P,\frac{3}{2}^-\rangle$&3090.2&3088&3054&2993&3033\\
$|1P,\frac{5}{2}^-\rangle$&3119.1&3092&3051&2947&3010\\
$|1D,\frac{1}{2}^+\rangle$&&3331&3287&3215&3354\\
$|1D,\frac{3}{2}^+\rangle$&&3322&3282&3231&3325\\
$|1D,\frac{3}{2}^+\rangle$&&3335&3298&3262&3335\\
$|1D,\frac{5}{2}^+\rangle$&&3298&3286&3173&3299\\
$|1D,\frac{5}{2}^+\rangle$&&3325&3297&3188&3308\\
$|1D,\frac{7}{2}^+\rangle$&&3296&3283&3136&3276\\
$|2S,\frac{1}{2}^+\rangle$&$3188$&3202&3088&3100&3164\\
$|2S,\frac{3}{2}^+\rangle$&&3237&3123&3126&3197\\
$|2P,\frac{1}{2}^-\rangle$&&3408&3384&3345&3427\\
$|2P,\frac{1}{2}^-\rangle$&&3446&3435&3359&3436\\
$|2P,\frac{3}{2}^-\rangle$&&3450&3415&3315&3408\\
$|2P,\frac{3}{2}^-\rangle$&&3461&3433&3330&3417\\
$|2P,\frac{5}{2}^-\rangle$&&3467&3427&3290&3393\\
\bottomrule[1pt]
\end{tabular}
\end{table}
Since lack of experimental data, the slope is not reliable enough to determine which model is most preferable. Reference \cite{Ebert:2011kk} satisfies Eq.~(\ref{eq:CF3}) (EFG), the slope is close to 1/2 and a universal description holds for theoretical data. As for two values of $n$, there are two figures obtained by the same group \cite{Shah:2016nxi,Shah:2016mig}, whose figures refered to STRV and STRV2 in Fig. \ref{omegacn12s}, respectively, are slightly different from each other. The second has larger slope than the first one and hence, it can be rejected from our point of view.
The other two diagrams plotted with the results from CL \cite{Chen:2017gnu} and EFG \cite{Ebert:2011kk} models are similar to the plot obtained from Ref. \cite{Shah:2016nxi} and it may be possible that these three could be close to experiments, which we expect to have in future.

\section{Conclusions and discussion}\label{sec5}

In this article, using the Regge-like formula $(M-m_Q)^2=\pi\sigma L$ of Eq.~(\ref{eq:CF3}) between hadron mass $M$ and angular momentum $L$, we have analyzed heavy-light systems, $D/D_s/B/B_s$ mesons and all the charmed and bottom baryons with a principal quantum number $n=1$, i.e., $\Lambda_Q/\Sigma_Q/\Xi_Q/\Xi'_Q/\Omega_Q$ with $Q=c, b$, in which only one heavy quark is included. We have adopted only the confirmed values cited in Ref. \cite{Agashe:2014kda} to plot figures.

{Light quarks, $u$, $d$, and $s$, form the so-called chiral particles, e.g., $\pi$, $K$, etc. These quarks have the current quark masses, i.e., very tiny masses. When these quarks are dressed with gluon clouds, they become constituent quarks. In our paper, we treat light quarks as current ones in the string picture so that their masses vanish in the chiral limit. Hence, we should subtract only heavy quark mass from the hadron. Light current quark masses should not be subtracted from the hadron because a part of a light current quark mass comes from gluon energy. We have numerically checked the cases in which a light current quark mass or a light current diquark mass is subtracted from the hadron. We have plotted figures, $L$ vs. $(M-m_Q-m_q)^2$ or  $(M-m_Q-m_{qq})^2$ and have found that line slopes become much smaller than 1/2 and universal description does not hold.}

For heavy-light mesons with the same $L$, we have used the average mass value of the $L$-wave states for a plot, i.e., the average value of ${}^{2S+1}L_J={}^1S_0$ and ${}^3S_1$ states for $S$-wave, the average value of ${}^1P_1$, ${}^3P_1$, ${}^3P_0$ and ${}^3P_2$ states for $P$-wave, etc.
For heavy-light baryons with the same $L$, we only have a singlet $J=1/2$ for $S$-wave, we have used the average value of $J=1/2$ and $3/2$ states for $P$-wave, the average value of $3/2$ and $5/2$ states for $D$-wave with $S_{qq}=0$ (a good type), etc. If there are isospin nonsinglet states, we, of course, have averaged over isospin states.

Numerical plots have been obtained for all the heavy-light mesons of experimental data whose slopes become nearly equal to 1/2 of that for light mesons as expected. A universal description also holds for all the heavy-light mesons, i.e., one unique line is enough to describe both of charmed and bottom heavy-light systems. Surprisingly enough it has been found that both of $D/B$ and $D_s/B_s$ have had almost the same unique lines as can be seen from Fig. \ref{DBmesons}. We have also checked the theoretical model of Ref. \cite{Ebert:2009ua} whether this model also obeys the above rules which are satisfied by experimental data and have found that the EFG model also supports their slopes being 1/2 and a universal description.
We have predicted the averaged mass values of $B$ and $B_s$ with $L=2$ as 6.01 and 6.13 GeV, respectively.

Regarding that charmed and bottom baryons consist of one heavy quark and one light cluster of two light quarks (diquark), we have applied the formula Eq.~(\ref{eq:CF3}) to all the heavy-light baryons including recently discovered $\Omega_c$'s and have found that experimental values of all the heavy-light baryons, $\Lambda_Q/\Sigma_c/\Xi_Q/\Xi'_Q/\Omega_Q$ with $Q=c, b$, if they exist, have well satisfied the formula, $(M-m_Q)^2=\pi\sigma L$ and the coefficient of $L$ is close to 1/2 compared with that for light hadrons.
Since there exist experimental data both for $\Lambda_c$ and $\Lambda_b$, we have seen univarsal description only for $\Lambda_Q$.
We have also checked the model calculations of Refs. \cite{Ebert:2011kk,Chen:2017gnu,Shah:2016mig,Shah:2016nxi} whether they satisfy Eq. (\ref{eq:CF3}), have slope close to 1/2, and a universal description holds and we have found that they really do satisfy all these rules except for Ref. \cite{Shah:2016nxi}.
Because of unknown assignments of five $\Omega_c$'s, we have also provided Fig. \ref{omegacn12s} with $n=1, 2$ which can be used for future analysis.

When looking at figures for a universal description, one notices that the slope for bottom heavy-light system has smaller value than that for charmed one. This is understandable because the heavy-light system with a $b$ quark is dominated by $m_b$ compared with that with $m_c$. The slope for bottom system is much closer to 1/2 than that for charmed one.

We have also predicted the average mass values of $\Lambda_b$, $\Sigma_c$, $\Xi_c$, and $\Omega_c$ with $L=2$ as 6.15, 3.05, 3.07, and 3.34 GeV, respectively, by using the straight lines for these baryons. These baryons should be averaged over two spin states $(3/2^+,5/2^+)$ for $\Lambda_b$ and $\Xi_c$ and over six spin states $(1/2^+,3/2^+,3/2^+,5/2^+,5/2^+,7/2^+)$ for $\Sigma_c$ and $\Omega_c$.
Our results of  all the heavy-light baryons suggest that heavy-light baryons can be safely regarded and treated as heavy quark-light cluster configuration.

Finally, we would like to comment on the reason why we could analyze all the heavy-light systems. This can be done because we have ignored $LS$ and $SS$ couplings which cause the small mass splittings among states with the same $L$. Otherwise we would have immediately faced serious problems, e.g., other than $\Lambda_Q$ and $\Xi_Q$ with $S_{qq}=0$ (good type), the heavy-light baryons with $S_{qq}=1$ (bad type) have combersom interactions as has been pointed out in Ref. \cite{Chen:2014nyo}. Hence, our way of analysis of heavy-light systems is the important and powerful tool to analyze experimental data as well as model calculations since by analyzing data, we can judge whether expeimental data observed are reliable and which model should be adopted or is reliable.

Nonlinearity of the parent Regge trajectoris is observed in model calculations of Ref. \cite{Ebert:2011jc}. Our analysis suggests that instead of mass, gluon flux energy should be used to relate it to angular momentum for heavy quark systems. Actually, nonlinearity of the parent Regge trajectory for $B_c$ obtained in Ref. \cite{Ebert:2011jc} can be remedied by adopting our formula Eq. (\ref{eq:ML}), which gives us a linear trajectory. That is, for heavy quark systems, heavy quark mass dominates to determine the curves in the ordinary Regge trajectories, while from our point of view, gluon flux energy determines behavior of heavy quark systems.

Future measurements of higher orbitally and radially excited states and their masses of heavy quark systems by LHCb and forthcoming BelleII are waited for to test our observation.


\section*{Acknowledgements}

This work is partly supported by the National
Natural Science Foundation of China under Grant
Nos. 11705056 and 11475192, as well as supported, in part, by the
DFG and the NSFC through funds provided to the Sino-German
CRC 110 Symmetries and the Emergence of Structure in QCD.
This work is also supported by China Postdoctoral Science Foundation under Grant No. 2016M601133.
T. Matsuki wishes to thank Yubing Dong of IHEP,
and Xiang Liu of Lanzhou university for their kind hospitality at each institue and university
where part of this work was carried out.

\end{document}